\documentclass
[aps,twocolumn,showpacs,tightenlines,superscriptaddress]{revtex4}
\usepackage{amsmath}
\usepackage{amsfonts}
\usepackage{amssymb}
\usepackage{graphicx}

\begin{document}
\title{Experimental Violation of Local Realism by Four-photon
Greenberger-Horne-Zeilinger Entanglement}
\author{Zhi Zhao}
\affiliation{Department of Modern Physics, University of Science
and Technology of China, Hefei, Anhui 230027, People's Republic of
China}
\author{Tao Yang}
\affiliation{Department of Modern Physics, University of Science
and Technology of China, Hefei, Anhui 230027, People's Republic of
China}
\author{Yu-Ao Chen}
\affiliation{Department of Modern Physics, University of Science
and Technology of China, Hefei, Anhui 230027, People's Republic of
China}
\author{An-Ning Zhang}
\affiliation{Department of Modern Physics, University of Science
and Technology of China, Hefei, Anhui 230027, People's Republic of
China}
\author{Marek \.Zukowski}
\affiliation{Instytut Fizyki Teoretycznej i Astrofizyki
Uniwersytet Gdanski, PL-80-952 Gdansk, Poland}
\author{Jian-Wei Pan}
\affiliation{Department of Modern Physics, University of Science
and Technology of China, Hefei, Anhui 230027, People's Republic of
China} \affiliation{Institut f\"{u}r Experimentalphysik,
Universitat Wien, Boltzmanngasse 5, 1090 Wien, Austria}


\begin{abstract}
We report the first experimental violation of local realism in
four-photon Greenberger-Horne-Zeilinger (GHZ) entanglement. In the
experiment, the non-statistical GHZ conflicts between quantum
mechanics and local realism are confirmed, within the experimental
accuracy, by four specific measurements of polarization
correlations between four photons. In addition, our experimental
results not only demonstrate a violation of
Mermin-Ardehali-Belinskii-Klyshko inequality by $76$ standard
deviations, but also for the first time provide sufficient
evidence to confirm the existence of genuine four-particle
entanglement.

\end{abstract}
\maketitle

Multi-particle entanglement not only plays a crucial role in
fundamental tests of quantum mechanics (QM) versus local realism
(LR), but is also at the basis of nearly all quantum information
protocols such as quantum communication and quantum computation
\cite{dirk00}. Since the seminal work of Greenberger, Horne,
Zeilinger (GHZ) \cite{zeilinger89}, the research on multi-particle
entanglement has received much attention. In contrast to the case
of two-particle entanglement where only statistical correlation
predicted by QM is inconsistent with LR, in the case of maximally
entangled states of more than two particles (i.e. the so-called
GHZ states) a conflict with LR arises even for nonstatistical
predictions of QM \cite{zeilinger89}. Further, QM can violate the
multi-particle Bell-type inequalities imposed by LR by an amount
that grows exponentially with the number of entangled particles
\cite{mermin90a,ardehali92,be93,zuk93}, that is, going to higher
entangled systems the conflict between QM and LR becomes ever
stronger.

In recent years, entanglement of three photons has been realized
experimentally \cite{dirk99} and used to demonstrate the extreme
GHZ contradiction between QM and LR \cite{pan00}. Meanwhile,
entanglement of three atoms \cite{rau00} or four ions
\cite{sack00} has also been demonstrated, yet in these two cases
the quality of the entangled states still needs to be improved
significantly in order to be useful for tests of LR and for
quantum information processing. Though significant experimental
progress has been achieved, all the above experiments suffer from
a loophole in confirming genuine multi-particle entanglement
\cite{see01}. This is due to the fact that the data measured in
any of the above $N$-particle entanglement experiments can be
explained by a hybrid model in which only less than $N$ particles
are entangled \cite{see01}. Using the highly pure four-photon
entanglement achieved in a recent experiment \cite{pan01}, it is,
in principle, possible to exclude such a hybrid model by showing a
sufficient violation of Bell-type inequalities. However, due to
the very low coincidence rate in the experiment it was not
possible to show such a violation. Therefore, the loophole problem
still remains unsolved.

In this Letter, we develop a high intensity source of four-photon
GHZ entanglement \cite{weinfurter03}, by which we report the first
four-observer test of GHZ contradiction, and for the first time
provide sufficient experimental evidence to confirm the existence
of genuine four-particle entanglement, hence closing the possible
loophole of a hybrid model.
\begin{figure}
[ptb]
\begin{center}
\includegraphics[height=2.5374in, width=2.9334in] {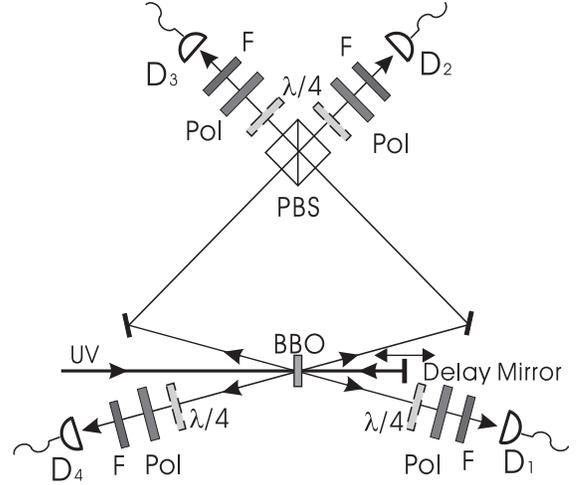}
\caption{Experimental setup for observing high intensity
four-photon GHZ entanglement. Two pairs of entangled photons are
produced by passing a UV laser pulse through a BBO crystal twice.
The UV laser with a central wavelength of 394nm has a pulse
duration of 200fs, a repetition rate of 76MHz and an average pump
power of 450mW. By optimizing the collection efficiency
\cite{weinfurter01}, we are able to observe about $2\times10^4$
entangled pairs per second for each pair behind 3.6nm filters (F)
of central wavelength 788nm. Coincidences between detectors $D_1$,
$D_2$, $D_3$ and $D_4$ exhibit four-photon GHZ entanglement.
Polarizers (POL) and quarter wave-plates ($\lambda/4$) are used to
perform measurement of linear $H^{\prime}/V^{\prime}$ or $R/L$
polarization.}
\end{center}
\end{figure}

To demonstrate the four-photon GHZ contradiction, we first
generate four-photon entanglement using the technique developed in
a previous experiment \cite{pan01}. As shown in Fig.1, a pulse of
ultraviolet (UV) light passes through a BBO crystal twice to
produce two polarization-entangled photon pairs. One photon out of
each pair is then steered to a polarization beam splitter (PBS)
where the path lengths of each photon have been adjusted such that
they arrive simultaneously. After the two photons pass through the
PBS, correlations due to four-photon GHZ entanglement,
\begin{equation}
\left\vert \Psi\right\rangle =\frac{1}{\sqrt{2}}(\left\vert
H\right\rangle _{1}\left\vert V\right\rangle _{2}\left\vert
V\right\rangle _{3}\left\vert H\right\rangle _{4}+\left\vert
V\right\rangle _{1}\left\vert H\right\rangle _{2}\left\vert
H\right\rangle _{3}\left\vert V\right\rangle _{4}), \label{state}
\end{equation}
can thus be observed, if all four detectors click (see
\cite{zeilinger97}). Here $H$ ($V$) denotes horizontal (vertical)
linear polarization.

The observed four-fold coincident rate of the desired component
$HVVH$ or $VHHV$ was about 1.3 per second, which is almost two
orders of magnitude brighter than the previous experiment
\cite{pan01}. The ratio between any of the desired four-fold
events $HVVH$ and $VHHV$ to any of the 14 other nondesired ones is
better than $60:1$. To confirm the state (\ref{state}) is indeed
in a coherent superposition, we have performed polarization
measurements on the four photons in the $H^{\prime}/V^{\prime}$
basis, where $\left\vert H^{\prime}\right\rangle =1/\sqrt{2}\left(
\left\vert H\right\rangle +\left\vert V\right\rangle \right)$ and
$\left\vert V^{\prime}\right\rangle =1/\sqrt{2}\left(  \left\vert
H\right\rangle -\left\vert V\right\rangle \right)$.  In Fig. 2, we
compare the count rates of
$H^{\prime}H^{\prime}H^{\prime}H^{\prime}$ and
$H^{\prime}H^{\prime}H^{\prime}V^{\prime}$ components as we move
the delay mirror. At zero delay, the latter component is
suppressed with a visibility of $0.84\pm0.03$, hence confirming
the coherent superposition of $HVVH$ and $VHHV$. The high
visibility and high intensity achieved indicates that the source
is good enough to demonstrate the GHZ contradiction between QM and
LR.
\begin{figure}
[ptb]
\begin{center}
\includegraphics[
natheight=3.173000in, natwidth=4.205600in, height=2.4076in,
width=3.1825in] {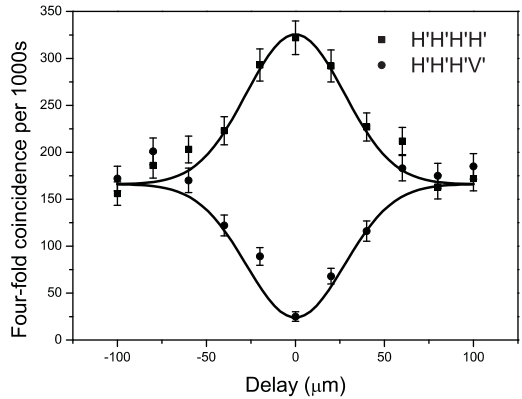} \caption{Typical experimental results
for polarization measurements on all four photons in the
$H^{\prime}/V^{\prime}$ basis. The coincidence rates of
$H^{\prime}H^{\prime}H^{\prime}H^{\prime}$ and
$H^{\prime}H^{\prime}H^{\prime}V^{\prime}$ components are shown as
a function of the pump delay mirror position. The high visibility
obtained at zero delay implies that four photons are indeed in a
coherent superposition.}
\end{center}
\end{figure}

Let us analyze the QM predictions for the four-photon state
(\ref{state}). Since the polarization states of a photon are a
realization of a qubit, one can represent $\vert H\rangle$ by
column vector $\left( \begin{array}{c} 0 \\ 1 \end{array}\right)$
and $\vert V\rangle$ by column vetcor $\left( \begin{array}{c} 1 \\
0 \end{array}\right)$. That is, they can be treated as the two
eigenvectors of Pauli operator $\sigma_{x}$ of eigenvalues $+1$
and $-1$ respectively. Adopting the methods of Refs.
\cite{mermin90,pan00}, we consider measurements of linear
polarization $H^{\prime}/V^{\prime}$, or of circular polarization
$R/L$, where $\left\vert R\right\rangle =\frac{1}{\sqrt{2}}\left(
\left\vert H\right\rangle +i\left\vert V\right\rangle \right)  $
and $\left\vert L\right\rangle =\frac{1}{\sqrt{2}}\left(
\left\vert H\right\rangle -i\left\vert V\right\rangle \right)  $
can be represented as the two eigenstates of Pauli operator
$\sigma_{y}$ with eigenvalues $\pm1$. We shall call a measurement
of $H^{\prime}/V^{\prime}$ linear polarization as a $\sigma_{x}$
measurement and one of $R/L$ circular polarization as a
$\sigma_{y}$ measurement.

To illustrate the GHZ conflict between QM and LR, we first
consider three specific measurements of polarization correlations
between four photons:
\begin{equation}
\begin{array}
[c]{ccc} \sigma_{x}\sigma_{x}\sigma_{x}\sigma_{x}, &
\sigma_{x}\sigma_{y}\sigma_{x}\sigma_{y}, & \sigma_{x}\sigma_{x}
\sigma_{y}\sigma_{y}
\end{array}
\label{basis}
\end{equation}
where, for example, $\sigma_{x}\sigma_{x}\sigma_{y}\sigma_{y}$
denotes a joint measurement of linear polarization
$H^{\prime}/V^{\prime}$ on photons 1 and 2, and circular
polarization $R/L$ on photons 3 and 4. The three operators in Eq.
(\ref{basis}) commute with each other and the state (\ref{state})
is their common eigenstate with the eigenvalue $+1$. Thus, in any
of the three measurements, the total number of photons that carry
either $V^{\prime}$ or $L$ polarization (i.e. with eigenvalue
$-1$) must be even. For example, in a
$\sigma_{x}\sigma_{x}\sigma_{y}\sigma_{y}$ measurement, only
polarization combinations $H^{\prime}H^{\prime}RR$,
$H^{\prime}H^{\prime}LL$, $H^{\prime}V^{\prime}RL$,
$H^{\prime}V^{\prime}LR$, $V^{\prime}H^{\prime}RL$,
$V^{\prime}H^{\prime}LR$, $V^{\prime}V^{\prime}RR$, and
$V^{\prime}V^{\prime}LL$ arise. Similar constraints can also be
obtained for the other two measurements of (2).
\begin{figure}
[ptb]
\begin{center}
\includegraphics[ height=4.1329in,
width=2.8193in] {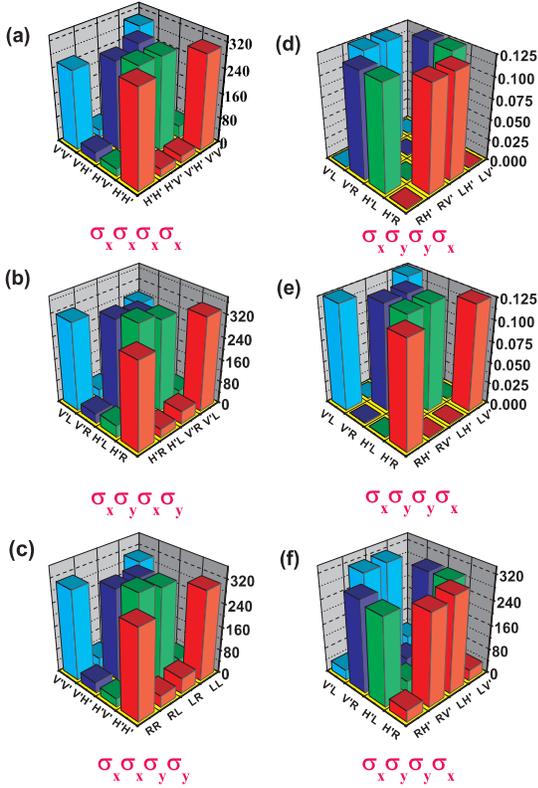} \caption{Experimental results observed
in the first three experiments (a)-(c), and predictions of QM and
of LR (normalized), and observed results for the
$\sigma_{x}\sigma_{y}\sigma_{y} \sigma_{x}$ measurement (d)-(f).
The visibilities in (a)-(c) are $0.820\pm0.011$, $0.807\pm0.011$
and $0.781\pm0.012$, respectively. The experimental results in (f)
are in agreement with the QM predictions (d) while in conflict
with LR (e), with a visibility of $0.789\pm0.012$. The integration
time of each four-fold coincidence is 1000s.}
\end{center}
\end{figure}

We now analyze what are the implications for LR. Although any
specific result obtained in any joint measurement on less than
four photons is maximally random, one can still presume that, each
photon carries Einstein-Podolsky-Rosen (EPR) elements of reality
\cite{EPR} for both $\sigma_{x}$ and $\sigma_{y}$ measurements
that determine the specific individual measurement result
\cite{note1}. This is because in every one of the three
measurements, any individual measurement result - both for
circular polarization and for linear polarization
$H^{\prime}/V^{\prime}$ - can be predicted with certainty for
every photon given the corresponding measurement results of the
other three \cite{pan00,mermin90}.

For any photon $i$ we call these elements of reality $X_{i}$ with
values +1(-1) for $H^{\prime}(V^{\prime})$ polarizations and
$Y_{i}$ with values +1(-1) for $R\left(  L\right)$; we thus obtain
the relations
$X_{1}X_{2}X_{3}X_{4}=X_{1}Y_{2}X_{3}Y_{4}=X_{1}X_{2}Y_{3}Y_{4}
=+1$, in order to be able to reproduce the quantum predictions on
all three measurements in Eq. (\ref{basis}). Furthermore,
according to LR, any specific measurement for $\sigma_{x}$ or
$\sigma_{y}$ must be independent of whether a $\sigma_{x}$ or
$\sigma_{y}$ measurement is performed on the other photons. As
$X_{i}X_{i}=+1$ and $Y_{i}Y_{i}=+1$, we can write
$X_{1}Y_{2}Y_{3}X_{4}=\left(  X_{1} X_{2}X_{3}X_{4}\right) \left(
X_{1}Y_{2}X_{3}Y_{4}\right)  \left( X_{1}X_{2}Y_{3}Y_{4}\right)$
and obtain $X_{1}Y_{2}Y_{3}X_{4}=+1$.

Therefore, the existence of the elements of reality implies that,
performing a $\sigma_{x}\sigma _{y}\sigma_{y}\sigma_{x}$
measurement on the state (\ref{state}), one should obtain the
product of the eigenvalues with $+1$. Thus, from a local realistic
point of view the only possible results for a
$\sigma_{x}\sigma_{y}\sigma_{y}\sigma_{x}$ measurement are
$H^{\prime}RRH^{\prime}$, $H^{\prime}RLV^{\prime}$,
$H^{\prime}LRV^{\prime}$, $H^{\prime}LLH^{\prime}$,
$V^{\prime}RRV^{\prime} $, $V^{\prime}RLH^{\prime}$,
$V^{\prime}LRH^{\prime},$ and $V^{\prime} LLV^{\prime}$ (as shown
in Fig. 3e).

However, according to QM, the state (\ref{state}) is an eigenstate
with eigenvalue -1 for operator $\sigma
_{x}\sigma_{y}\sigma_{y}\sigma_{x}$. Thus, QM predicts that the
only possible results for a
$\sigma_{x}\sigma_{y}\sigma_{y}\sigma_{x}$ measurement are
$H^{\prime}RRV^{\prime}$, $H^{\prime}RLH^{\prime}$,
$H^{\prime}LRH^{\prime}$, $H^{\prime}LLV^{\prime}$,
$V^{\prime}RRH^{\prime}$, $V^{\prime}RLV^{\prime}$,
$V^{\prime}LRV^{\prime}$, and $V^{\prime}LLH^{\prime}$ (as shown
in Fig. 3d). Thus we conclude that the predictions by LR is
completely opposite to the predictions by QM and vice versa. It is
the GHZ contradiction between LR and QM that exhibits a more
powerful refutation of the existence of elements of reality than
the one provided by Bell's theorem for two-particle entanglement.

To demonstrate experimentally this conflict between LR and QM, we
perform four polarization correlation measurements on the four
spatially separated photons. The observed results for the first
three measurements are shown in Figs. 3a, 3b, 3c. Each measurement
consists of 16 possible outcomes and ideally only eight of them
should occur. However, since in reality no experiment can ever be
perfect, even the outcomes which should not occur will occur with
some small probabilities. Thus, if we are allowed to assume that
the spurious events are attributable to the unavoidable
experimental errors, then within the experimental accuracy we can
conclude that the desired correlations in the three measurements
confirm the quantum predictions for our GHZ entanglement.

In Figs. 3d, 3e, 3f , we compare the predictions of QM and LR with
the results of the fourth
$\sigma_{x}\sigma_{y}\sigma_{y}\sigma_{x}$ measurement. The
results show that, within experimental error, the four-fold
coincidences predicted by QM occur, and not those predicted by LR.
In this sense, we claim that we have experimentally realized the
first four-particle test of local realism following the GHZ
argument. For the purists, we may note that there is a derivation
of the GHZ paradox for situations involving up to $25\%$ (data
flipping) error rate \cite{RYFF}, that is for rates much higher
than observed in the experiment ($\sim 11\%$).

The conflict between the quantum predictions for the GHZ states
and local realism can also be shown via violation of a suitable
Bell inequality. In this case taking account of the errors is
straightforward. A number of inequalities for $N$-particle GHZ
states have been derived \cite{mermin90a,ardehali92,be93,zuk93}.
According to the optimal MABK inequality for four-particle GHZ
state \cite{ardehali92}, LR imposes a constraint on statistical
correlations of polarization measurements on the four-photon
system as the following:
\begin{equation}
\left\vert \left\langle A\right\rangle \right\vert
\leq2,\label{inequality}
\end{equation}
where
\begin{align}
A & =
\frac{1}{2}(\sigma_{x}\sigma_{x}\sigma_{x}-\sigma_{x}\sigma_{y}\sigma_{y}
+\sigma_{y}\sigma_{x}\sigma_{y}+\sigma_{y}\sigma_{y}\sigma_{x})(\sigma_{a}+\sigma_{b})\nonumber\\
& +
\frac{1}{2}(\sigma_{y}\sigma_{y}\sigma_{y}-\sigma_{x}\sigma_{y}\sigma_{x}
+\sigma_{x}\sigma_{x}\sigma_{y}+\sigma_{y}\sigma_{x}\sigma_{x})(\sigma_{a}-\sigma_{b})
\end{align}
and $\sigma_{a}=\frac{1}{\sqrt{2}}(\sigma_{x}+\sigma_{y})$,
$\sigma_{b}=\frac{1}{\sqrt{2}}(\sigma_{x}-\sigma_{y})$, and they
correspond to measurements of two (othogonal) pairs of elliptic
polarizations. In Eq. (\ref{inequality}), for example,
$\left\langle \sigma_{x}\sigma_{x}\sigma_{x}\sigma_{a}
\right\rangle $ denotes the expectation value of a
$\sigma_{x}\sigma_{x}\sigma_{x}\sigma_{a}$ measurement on the four
photons. QM predicts a maximal violation of the constraint by a
factor of $2\sqrt{2}$. For the prefect quantum prediction the
visibility of the correlations can be reduced to as little as
$35.4\%$. Interestingly, a different set of measurements, than
those for the GHZ contradiction, are optimal in the case of this
inequality. Further, one could note that the inequalities derived
in \cite{zuk93} require only a visibility of $32.9\%$. All this
should be contrasted with the visibility consistent with the
result of ref. \cite{RYFF}, concerning the GHZ contradiction,
which is $50\%$. Therefore in order to get maximal possible
disagreement with LR, we performed another set of measurements.

To measure the expectation value of $A$, we need to perform
sixteen specific measurements such as
$\sigma_{x}\sigma_{x}\sigma_{x}\sigma_{a}, ...,
\sigma_{y}\sigma_{x}\sigma_{x}\sigma_{b}$. A $\sigma_a$
measurement on photon 4 is obtained if we insert in its path a
quarter wave plate (QWP), whose optical axis is set at
$45^{\circ}$ with respect to the horizontal direction. Then, the
two eigenstates of operator $\sigma_a$ are converted into linear
polarizations which are polarized along the directions of
$-22.5^{\circ}$ and $67.5^{\circ}$. In the same way, the two
eigenstates of operator $\sigma_b$ can be converted into
$-67.5^{\circ}$ and $22.5^{\circ}$ linear polarizations. The
average visibility observed in the experiment for the state
(\ref{state}) is $78.4\%$ and thus greatly exceed the minimum of
$35.4\%$. Substituting the experimental results into the left-hand
side of inequality (\ref{inequality}) gives
\begin{equation}
\left\vert \left\langle A\right\rangle \right\vert =4.433\pm
0.032,\label{value}
\end{equation}
which violate the inequality (\ref{inequality}) by over $76$
standard deviations, hence demonstrating the conflict between QM
and LR in four-photon GHZ entanglement.

Furthermore, the high visibilities also confirm the existence of
genuine four-photon entanglement in our experiment. To demonstrate
a full four-photon entanglement, two sufficient conditions, i.e.
the inequality $\left\vert \left\langle A\right\rangle
\right\vert>4$ and the so-called state preparation fidelity
$F(\rho)>1/2$, must be satisfied \cite{gisin98,see01}. Here the
state preparation fidelity is defined as
\begin{align}
F(\rho) &=\left\langle \Psi \right| \rho \left| \Psi
\right\rangle\nonumber\\
&=\frac 12\left( \left\langle HVVH\right| \rho \left|
HVVH\right\rangle\right. \left.+ \left\langle VHHV\right| \rho
\left| VHHV\right\rangle \right) \nonumber\\ &+ Re\left\langle
HVVH\right| \rho \left| VHHV\right\rangle
\end{align}
and for any state $\rho$, there is a simple identity:
\begin{equation}
\left| \left\langle A\right\rangle \right| = 8 \sqrt{2}
\mathop{\rm Re} \left\langle HVVH\left| \rho \right|
VHHV\right\rangle.
\end{equation}

Not only does the experimental result in Eq. (\ref{value})
significantly violate the inequality $\left\vert \left\langle
A\right\rangle \right\vert >4$, together with the observed
fractions of the desired components and the nondesired ones in the
$H/V$ basis it also gives $F(\rho)=0.840\pm0.007$, which is well
above the threshold of $1/2$. Thus, our experiment for the first
time provides unambiguous evidence for a full test of
four-particle entanglement, which excludes any hybrid
hidden-variable model to explain our experimental data.

In conclusion, we have demonstrated the statistical and
nonstatistical conflicts between QM and LR in four-photon GHZ
entanglement. However, it is worth noting that, as for all
existing photonic tests of LR, we also had to invoke the fair
sampling hypothesis due to the very low detection efficiency in
our experiment. Possible future experiments could include further
study of GHZ correlations over large distances with space-like
separated randomly switched measurements \cite{weihs}. Our work,
besides its significance in quantum foundations, could also be
applied to investigate the basic elements of quantum computation
with linear optics \cite{knill} and implement multi-photon quantum
secrete sharing \cite{zukowski}.

This work was supported by the National Natural Science Foundation
of China, Chinese Academy of Sciences and the National Fundamental
Research Program (under Grant No. 2001CB309303). MZ is supported
by the UG grant BW-5400-5-0256-3.


\begin{references}
\bibitem{note1} Note that, to argue, that photon 1 also carries element of
reality for a $\sigma_y$ measurement, we could consider a
$\sigma_{y}\sigma_{y}\sigma_{y}\sigma_{y}$ measurement for which
the state (1) is again an eigenstate with eigenvalue $+1$.
\end{references}
\end{document}